# Automated Hardware Logic Obfuscation Framework Using GPT

Banafsheh Saber Latibari[†], Sujan Ghimire[§], Muhtasim Alam Chowdhury[§], Najmeh Nazari[†],
Kevin Immanuel Gubbi[†], Houman Homayoun[†], Avesta Sasan[†], Soheil Salehi[§]
{bsaberlatibari, nnazari, kgubbi, hhomayoun, asasan}@ucdavis.edu
{sghimire, mmc7, ssalehi}@arizona.edu
[†]University of California Davis, [§]University of Arizona

*Abstract*—Obfuscation stands as a promising solution for safeguarding hardware intellectual property (IP) against a spectrum of threats including reverse engineering, IP piracy, and tampering. In this paper, we introduce Obfus-chat, a novel framework leveraging Generative Pre-trained Transformer (GPT) models to automate the obfuscation process. The proposed framework accepts hardware design netlists and key sizes as inputs, and autonomously generates obfuscated code tailored to enhance security. To evaluate the effectiveness of our approach, we employ the Trust-Hub Obfuscation Benchmark for comparative analysis. We employed SAT attacks to assess the security of the design, along with functional verification procedures to ensure that the obfuscated design remains consistent with the original. Our results demonstrate the efficacy and efficiency of the proposed framework in fortifying hardware IP against potential threats, thus providing a valuable contribution to the field of hardware security.

## I. INTRODUCTION

The application of Large Language Models (LLMs) has permeated diverse domains, revolutionizing fields beyond natural language processing (NLP) and extending into realms such as computer vision, drug discovery, code generation, and more. In recent years, LLMs like GPT have showcased remarkable adaptability and efficacy across various domains, showcasing their potential to transcend traditional boundaries and drive innovation in unexpected arenas like hardware design [3], [9], [10], [13].

The globalization of manufacturing processes and electronic hardware supply chains has been propelled by the imperative to enhance profitability while mitigating risk in a progressively advanced silicon industry. Nevertheless, numerous security features of hardware IPs have been compromised due to the increasing prevalence of successful hardware attacks [4]. Logic obfuscation techniques play a pivotal role in safeguarding intellectual property, protecting sensitive information, and thwarting unauthorized access or reverse engineering attempts in integrated circuits (ICs) and hardware systems. Automating the process of logic obfuscation and providing a framework for designing obfuscated IPs can significantly expedite the development cycle. While some research endeavors have explored the use of Machine Learning (ML) techniques [2] for this purpose, there remains a notable gap in the literature regarding the utilization of LLMs for automating logic obfuscation. Integrating LLMs into the logic obfuscation framework holds the potential to revolutionize the design process, enabling designers to rapidly prototype and deploy obfuscated IPs with minimal manual intervention and significantly expedite the development cycle.

In this paper, we present the following contributions:
- We introduce the first framework for logic obfuscation based on LLMs.
- We demonstrate the efficacy of our framework through experimentation, employing SAT attack and TrustHub benchmarks as metrics for evaluation.

## II. BACKGROUND

### A. Logic Locking Basics

Logic locking involves enhancing post-fabrication programmability by introducing supplementary gates termed key-programmable gates (key gates). These gates are manipulated by a secret element known as the key, integral to the logic locking mechanism [7]. Locking is implemented on a specific subset of internal nets within a design. Variations exist among existing solutions regarding the selection of locking gates, incorporation of dummy functions, and other transformations applied to each designated net [2]. In the next section, we delve into the details regarding the choice of existing approaches.

**Different Types of Logic Gates and Dummy Functions-** At the core of logic locking lies the deployment of specialized gates known as locking key gates, typically XOR/XNOR gates. These gates feature one input dedicated to the key signal, while the other input is linked to the net targeted for locking. Furthermore, the locking mechanism employs multiplexers and switch-boxes, where one input of the multiplexer represents the original function, while the other input is assigned a dummy function. The dummy function can take various forms, such as a constant, a primary input, a net from another fan-in cone in the design, or a newly generated random function. The select signal of the multiplexer serves as the key, determining the correct value associated with the original fan-in cone of the circuit. Additionally, the utilization of Look-Up Tables (LUTs) and Maxes enhances the locking capabilities, providing more robust security measures.

**Different Net Selection Methodologies-** Several net selection processes have been introduced in the literature. These include Random Insertion, where key gates are inserted without prior analysis of the circuit; Secure Logic Locking, which implements a set of heuristics to mitigate key-gate masking issues; Logic Cone Size, which aims to maximize the corruption caused by key gates; and Controllability and Observability-Based Insertion, where the selection process analyzes each net based on switching activity or the probability of affecting observed outputs.

*B. ML-Guided Logic-Locking*

Over the past decade, the field of logic locking and obfuscation has experienced substantial growth, marked by numerous contributions to both locking techniques and the methods used to attack them. However, there remains a clear need for a comprehensive framework that can effectively implement robust logic-locking solutions while accounting for a diverse array of available attacks and their corresponding countermeasures. Additionally, such a framework must possess flexibility to adapt to future attack strategies. This is where the necessity for a machine learning-based approach arises. Alagh et al. [2] introduced LeGO a learning-guided framework for enhancing the security of hardware IP through iterative obfuscation techniques. The framework employs simple key-based locking to the target IP at first. Following that the framework enters an iterative mechanism guided by feedback, gradually strengthening the IP against a predefined set of attacks. Hassan et al. [6] proposed SATConda, as a solution to the challenges posed by logic obfuscation, particularly against SAT attacks and their variants. Unlike traditional defense mechanisms which often incur significant power, performance, and area overheads, SATConda presents a neural network (NN)-based SAT-hard clause translator. This approach minimizes area and power overhead while maintaining the original functionality and enhancing security. SATConda operates with a SAT-hard clause generator that transforms existing conjunctive normal form (CNF) through subtle perturbations, such as the addition of inverters, buffers, or lightweight SAT-hard blocks. To facilitate efficient SAT-hard clause generation, SATConda employs a multilayer NN capable of learning feature dependencies, followed by a long short-term memory (LSTM) network for validating and backpropagating SAT-hardness, thus ensuring robust learning and translation processes.

*C. LLM in Hardware Design and Security*

**LLM in Hardware Design-** Hardware designs typically originate as specifications written in natural language, although they are ultimately expressed in Hardware Description Languages (HDLs). However, translating these specifications into the appropriate HDLs is a task typically performed by hardware engineers. This process is known to be both time-consuming and prone to errors. This prompts the exploration of AI or ML-based tools for translating specifications into HDL, with LLM emerging as a promising candidate for this task. The authors proposed chip-chat to investigate the potential advantages and obstacles associated with integrating LLMs into the HDL development process. Throughout the creation process of a novel 8-bit processor, they facilitate a directed yet open-ended "free chat" session, leveraging an LLM as a collaborative hardware architect [3]. Thakur et al. proposed VeriGen [13], which harnesses a vast repository of open-source Verilog code to train LLMs. In this method, five pre-trained LLM models, with parameters ranging from 345M to 16B, are fine-tuned for generating Verilog code. They are then rigorously evaluated using a diverse set of Verilog coding problems, along with corresponding test benches, to ensure functional correctness. A thorough comparison is undertaken against leading general-purpose LLMs, such as ChatGPT variants and PALM2. Researchers from NVIDIA proposed VerilogEval [9] a benchmarking framework customized to assess the performance of LLM in the domain of generating Verilog code for hardware design and verification. They provided a comprehensive evaluation dataset consisting of 156 problems obtained from HDLBits. RTLLM, an open-source benchmark for generating design RTL with natural language instructions, was proposed in this work [10]. They systematically evaluated the auto-generated design RTL by summarizing three progressive goals: syntax goal, functionality goal, and design quality goal. This benchmark could automatically provide a quantitative evaluation of any given LLM-based solution. Additionally, they introduced a prompt engineering technique named self-planning, to boost the performance of GPT-3.5. In this paper [5], they proposed a formal verification methodology for Design Under Verification (DUV), utilizing Z3 in the design. Alongside invariants, they employed mutation testing with LLM to ensure property quality. The methodology relied on OpenAI's GPT-4 to auto-generate formal properties from Verilog models. Experiments were conducted on ISCAS-85 C432 27-channel interrupt controller due to its complexity.

**LLM in Hardware Security-** The widespread adoption of system-on-chip (SoC) technology across diverse computing devices has heightened concerns about hardware security, notably the threat posed by hardware Trojans (HTs). SoCs, employed in automotive electronics, home automation systems, industrial automation systems, and medical devices, manage critical user data. However, their complex architecture, often comprising various IP cores, creates vulnerabilities. HTs, stealthily inserted during design or fabrication, can compromise system integrity upon activation, leading to data breaches and unauthorized access. Detecting and mitigating HTs, exacerbated by third-party IPs and intricate SoC components, present significant challenges in ensuring hardware security. Saha et al. [12] explore four distinct security tasks using LLMs. Firstly, they demonstrate the capability of LLMs to introduce potential vulnerabilities into RTL designs based on natural language descriptions provided by a well-crafted prompt. Subsequently, leveraging LLMs, they conduct a comprehensive evaluation of the security landscape in hardware designs, identifying vulnerabilities, weaknesses, and potential threats. Additionally, they examine LLMs' ability to identify coding issues that may lead to security bugs. Moreover, using LLMs, they verify whether designs adhere to specific security rules or policies and assess LLMs' proficiency in calculating security metrics, understanding security properties, and generating functional testbenches to identify weaknesses. Lastly, they investigate the effectiveness of LLMs in mitigating existing vulnerabilities within designs. Kokolakis et al. [8] explored the potential of LLMs in the offensive hardware security domain, specifically examining their assistance to attackers in inserting HTs into complex designs like CPUs. They tested a general-purpose LLM's ability to correlate system-level security concepts with specific module abstractions in hardware designs, overcoming context length limitations. By analyzing reduced code bases and instructing the LLM to insert trojan functionalities, they demonstrated a streamlined approach to HT insertion. To showcase their automated LLM-based HT insertion flow, they crafted a realistic HT for a modern RISC-V micro-architecture, testing its functionality on an FPGA board to target the integrity and availability of the RISCV CPU. This illustrated how LLMs can guide attackers through intricate hardware designs and facilitate the implementation of HT attacks. Meng et al. [11] discuss the potential of LLMs for hardware security assurance. This paper exhibits the development of a novel framework called NSPG, which stands for Natural Language Processing (NLP)-based Security Property Generator. The paper proposes an automated method for extracting security properties from hardware documentation using a specialized language model, HS-BERT. The motivation to generate this research is to address the issue of generating security properties for System-on-Chips (SoCs), which is crucial for validating the security of these systems. The NSPG framework is evaluated using OpenTitan SoC documentation and proves itself to successfully identify security properties, demonstrating the potential of LLMs in hardware security validation.

TABLE I
SUMMARY OF PREVIOUS STUDIES UTILIZING LLM IN VARIOUS DOMAINS OF HARDWARE DESIGNS.

| Name | Reference | Year | Category | Code is Available |
|---|---|---|---|---|
| Chip-chat | 3 | 2023 | Verilog and RTL File Generation | Yes |
| LLM-guided Methodology | 5 | 2024 | Hardware Design Verification | No |
| LLMs for HT Design | 8 | 2024 | Insertion of Hardware Trojans | No |
| VerilogEval | 9 | 2023 | Verilog Code Generation | Yes |
| RTLLM | 10 | 2023 | RTL File Generation | Yes |
| NSPG | 11 | 2023 | SoC Security Assurance | No |
| LLM for SoC Security | 12 | 2023 | SoC Security Verification | No |
| VeriGen | 13 | 2023 | Verilog Code Generation | Yes |

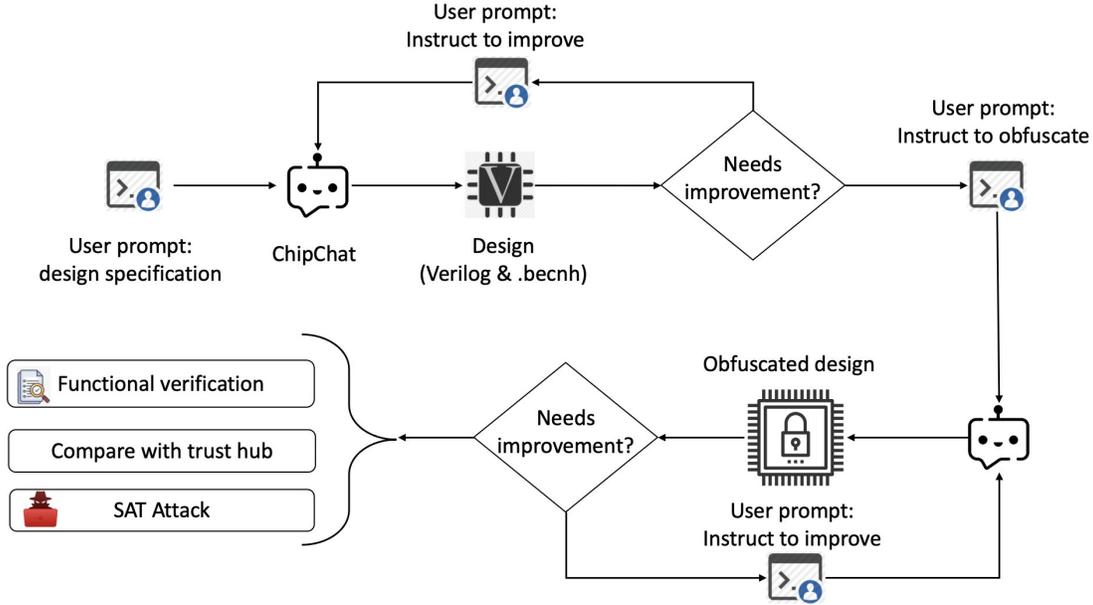

Fig. 1. Overview of the framework.

Table I provides a summary of previous studies utilizing LLM in various domains of hardware design. Notably, none of the previous works have employed LLM for logic obfuscation. These studies can be categorized into three main areas: HDL generation, design validation, and hardware trojan detection in the context of security.

## III. OBFUS-CHAT: GPT-BASED LOGIC OBFUSCATION FRAMEWORK

As mentioned in the preceding section, there has been no prior research focusing on the utilization of LLM for logic obfuscation. In this section, we delve deeper into our proposed

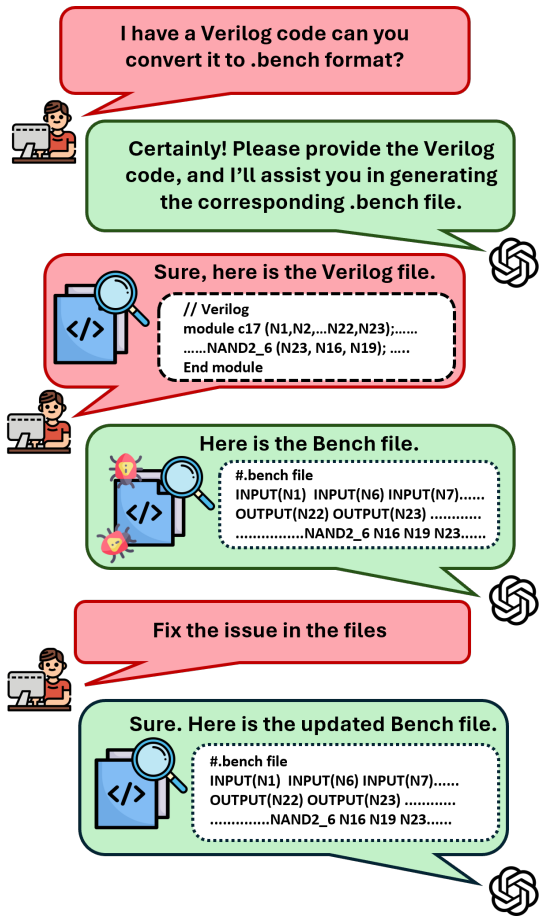

Fig. 2. Prompting for .bench format generation form verilog code.

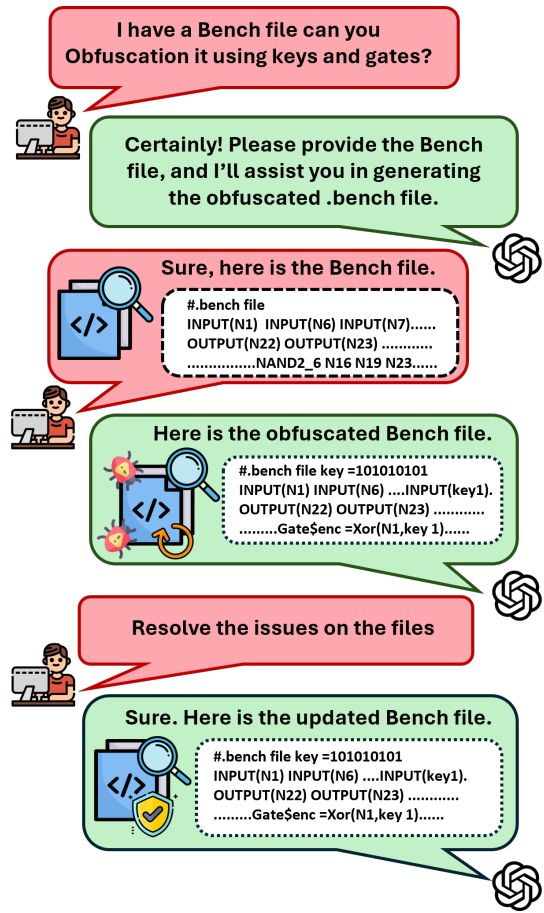

Fig. 3. Prompting for obfuscating the .bench file.

LLM-based logic obfuscation framework, known as obfuschat. Fig. 1 shows the overview of the framework. In the proposed framework, users utilize prompts to generate the Verilog implementation of the design. The generated Verilog code is then converted into a .bench file, followed by the obfuscation of the design. We offer GPT a range of obfuscation options, employing various key gates and nets. Moreover, we task GPT with generating an SAT-hard design based on these strategies:

- **Fan-out and Fan-in:** Introduce gates with high fan-out and fan-in. High fan-out gates increase the branching factor of the search space, while high fan-in gates complicate the dependencies between variables.
- **XOR Gates:** Incorporate XOR gates, which can introduce complexity and non-linearity to the design. XOR gates are known for their ability to break symmetries and create challenging SAT instances.
- **Complex Logic Structures:** Use complex logic structures such as multiplexers. These structures introduce intricate relationships between variables and constraints, making the SAT problem more difficult to solve.
- **Randomness:** Introduce randomness or pseudo-randomness into the design to prevent SAT solvers from exploiting predictable patterns or regularities.
- **Gate Placement:** Place gates strategically to maximize the difficulty of the SAT problem. Consider the trade-offs between gate placement, complexity, and performance.

The obfuscated .bench file generated is evaluated to ensure consistency with the original design, including having the same number of inputs, outputs, and functional behavior. Additionally, its SAT hardness is assessed by comparing the time required for a SAT attack on the obfuscated .bench file using our framework with that of an available obfuscated file in Trust Hub.

## IV. EXPERIMENTS

This section presents the results of our proposed framework. For simplicity, we have omitted the Verilog generation step in this stage and plan to incorporate it into future work. This will involve exploring various proposed HDL generation frameworks and integrating them with our design. At this stage, we have only tested combinational circuit forms from the ISCAS 85 benchmark [1]. We plan to address sequential circuits in our future work. We provide the Verilog code of the design to LLM, instructing them to generate the .bench file. Fig 2 shows our prompting approach to generate .bench file. We have tested GPT 3.5 it can handle small circuits like full adder and c17 but for larger designs, it returns the correct format but the file is not complete because of token limitation. For example for c432 we have used this *"Continue from the last line"* prompt several times to get the whole bench file. The generated output was based on only using 2 input gates. So, another prompt that was useful in generating the correct file is this: *"This is the Verilog code of c432 use the exact type of gates that are used in the design. For example, if it uses "NOT" you should use the same thing. If it uses 4 input "NAND" gate you should use the same thing."*

Figure 3 illustrates our prompting strategy for generating the obfuscated circuit. The process begins by receiving the .bench file and subsequently processing it for obfuscation. We tasked the GPT with specifying the number of key bits and the type of key gates, instructing it to obfuscate the design by randomly inserting key gates while preserving the circuit connectivity. The SAT attack successfully understands the obfuscated .bench file, enabling us to execute the attack. For larger designs, GPT encounters greater difficulty in retaining prompt information and preserving circuit connectivity. We plan to optimize this in our future work.

## V. Conclusion

In this study, we introduced a novel logic obfuscation framework based on LLMs, aiming to streamline the automation of the logic obfuscation process. We successfully tested our framework on combinational circuits, demonstrating its effectiveness in generating obfuscated designs. Looking ahead, our future work will extend to sequential circuits, exploring additional strategies such as scan chain and cycle insertion to further enhance the obfuscation process. Additionally, we plan to integrate our approach with existing HDL generation mechanisms, optimizing the entire framework for broader accessibility and applicability. Ultimately, our goal is to make the framework readily available for use by researchers and practitioners in the field.